\documentclass[nofootinbib,prd,twocolumn,showpacs,showkeys,preprintnumbers]{revtex4-1}
\usepackage{hyperref,amssymb,amsmath,mathrsfs,bm,graphicx}
\usepackage[utf8]{inputenc}
\usepackage{enumitem}
\input{epsf}
\begin{document}
\title {Quasi--hyperbolically symmetric $\gamma$-metric}
\author{L. Herrera}
\email{lherrera@usal.es}
\affiliation{Instituto Universitario de F\'isica
Fundamental y Matem\'aticas, Universidad de Salamanca, Salamanca 37007, Spain}
\author{A. Di Prisco}
\email{alicia.diprisco@ucv.ve}
\affiliation{Escuela de F\'\i sica, Facultad de Ciencias, Universidad Central de Venezuela, Caracas 1050, Venezuela}
\author{J. Ospino}
\email{j.ospino@usal.es}
\affiliation{Departamento de Matem\'atica Aplicada and Instituto Universitario de F\'isica
Fundamental y Matem\'aticas, Universidad de Salamanca, Salamanca 37007, Spain}
\author{J.  Carot}
\email{jcarot@uib.cat}
\affiliation{Departament de  F\'{\i}sica, Universitat Illes Balears, E-07122 Palma de Mallorca, Spain}

\date{\today}
\begin{abstract}
We carry out a systematic study on the motion of test particles in the region inner to the naked singularity of a quasi--hyperbolically symmetric $\gamma$-metric. The geodesic equations are written and analyzed in detail. The obtained results are contrasted with the corresponding results obtained for the axially symmetric $\gamma$-metric, and the hyperbolically symmetric  black hole. As in this latter case, it is found that test particles experience a repulsive force within the  horizon (naked singularity), which prevents them to reach the center. However in the present case this behavior is affected by the parameter $\gamma$ which measures the departure from the hyperbolical symmetry. These results are obtained for radially moving particles as well as for particles moving in the $\theta-r$ subspace. Possible relevance of these  results in the explanation of extragalactic jets, is brought out. 
\end{abstract}
\date{\today}
\maketitle
\section{Introduction}
In a recent  paper  \cite{1} an alternative  global description of the  Schwarzschild black hole  has been proposed. The motivation behind such an endeavor  was, on the one hand    the fact that the space--time within the horizon, in the classical picture,  is necessarily non--static or, in other words, that  any transformation that maintains the static form of the Schwarzschild metric (in the whole space--time) is unable to remove the coordinate singularity appearing on the horizon in the line element  \cite{rosen}.
Indeed, as  is well known,  no static observers can be defined inside the horizon (see \cite{Rin,Caroll} for a discussion on this point). This conclusion becomes intelligible if we recall  that the Schwarzschild horizon is also a Killing horizon, implying that the time--like Killing vector existing outside the horizon, becomes space--like inside it.

On the other hand, based on the physically reasonable point of view that  any equilibrium final state of a physical process should  be static, it would be desirable  to have  a static solution over the whole space--time.

Based on the arguments above, the following model was proposed in \cite{1}.

Outside the horizon ($R >2M$) one has the usual Schwarzschild  line element corresponding to the spherically symmetric vacuum solution to the Einstein  equations, which in polar coordinate reads (with signature $+2$)
\begin{eqnarray}
ds^2&=&-\left(1-\frac{2M}{R}\right)dt^2+\frac{dR^2}{\left(1-\frac{2M}{R}\right)}+R^2d\Omega^2, \nonumber \\ d\Omega^2&=&d\theta^2+\sin^2 \theta d\phi^2.
\label{w2}
\end{eqnarray}

This metric is static and spherically symmetric, meaning that it admits four Killing vectors:
\begin{eqnarray}
\mathbf{\xi}_{(\mathbf{0})} = \partial _{\mathbf{t}}, \quad {\bf \xi_{(2)}}=-\cos \phi \partial_{\theta}+\cot\theta \sin\phi \partial_{\phi},\nonumber \\
{\bf \xi_{(1)}}=\partial_{\phi}, \quad {\bf \xi_{(3)}}=\sin \phi \partial_{\theta}+\cot\theta \cos\phi \partial_{\phi}.
\label{2cmh}
\end{eqnarray}

The solution proposed for $R <2 M$  (with signature $-2$) is
\begin{eqnarray}
ds^2&=&\left(\frac{2M}{R}-1\right)dt^2-\frac{dR^2}{\left(\frac{2M}{R}-1\right)}-R^2d\Omega^2, \nonumber \\ d\Omega^2&=&d\theta^2+\sinh^2 \theta d\phi^2.
\label{w3}
\end{eqnarray}

This is a static solution, meaning that it admits the time--like Killing vector  $\mathbf{\xi }_{(\mathbf{0})}$, however  unlike (\ref{w2})  it is not spherically symmetric, but hyperbolically symmetric, meaning that it admits  the three Killing vectors
\begin{eqnarray}
{\bf \chi_{(2)}}=-\cos \phi \partial_{\theta}+\coth\theta \sin\phi \partial_{\phi},\nonumber \\
{\bf \chi_{(1)}}=\partial_{\phi}, \quad {\bf \chi_{(3)}}=\sin \phi \partial_{\theta}+\coth\theta \cos\phi \partial_{\phi}.
\label{2cmhy}
\end{eqnarray}

Thus  if one wishes to keep sphericity within the horizon, one should abandon staticity, and   if one wishes to keep staticity within the horizon, one should abandon sphericity.

The classical picture of the black hole entails sphericity within the horizon, instead in \cite{1} we have proceeded differently and have assumed staticity within the horizon.

The three Killing vectors (\ref{2cmhy})  define the hyperbolical symmetry.
Space--times endowed with hyperbolical symmetry have  previously  been the subject of research  in different contexts (see \cite{Ha}--\cite{hn5} and references therein).

In \cite{2nc} a general study of  geodesics in the spacetime described by (\ref{w3}) was presented (see also \cite{Lim}), leading to some interesting conclusions about the behavior of  a test   particle in this new picture of the Schwarzschild  black hole, namely:
\begin{itemize}
\item the gravitational force inside the region $R<2M$ is repulsive.
\item test particles cannot reach the center.
\item test particles can cross the horizon outward, but only along the $\theta=0$ axis.
\end{itemize}

These intriguing results reinforces further the interest on this kind of systems.

The procedure used in \cite{1} to obtain (\ref{w3}) may be used to obtain  hyperbolic versions of other spacetimes. Of course in this case the obtained metric may not admit  all the Killing vectors describing the hyperbolical symmetry (\ref{2cmhy}), and it will not describe a black hole but a naked singularity. We shall refer to these space--times as quasi--hyperbolical.

It is the purpose of this work  to delve deeper into this issue, by considering a specific  quasi-- hyperbolical space--time. Thus we shall analyze the quasi--hyperbolical  version of the $\gamma$-metric \cite{Bach, Vor, WE, Duncan}. In particular we endeavor to analyze the geodesic structure of this space--time, and  to  contrast it with the corresponding geodesics of the hyperbolically symmetric version of the Schwarzschild metric discussed  in \cite{2nc} and with the geodesic structure of the $\gamma$-metric discussed in \cite{gammas}.

The motivation for this choice is twofold, on the one hand the $\gamma$-metric corresponds to a solution of the Laplace equation, in cylindrical
coordinates, with the same Newtonian source image \cite{Bonnor} as the
Schwarzschild metric (a rod). On the other hand,  it has been proved  \cite{Herrera1} that by extending the length of the rod
to infinity one obtains the Levi--Civita spacetime. At the same time a link
was established between the parameter $\gamma$, measuring the mass density of
the rod in the $\gamma$-metric, and the parameter $\sigma$, which is thought to be
related to the  energy density of the source of the Levi--Civita
spacetime. The limit of the $\gamma$-metric when extending its rod source image to an infinite
length produces, intriguingly, the flat Rindler spacetime. This result enhances even more the peculiar character of the
$\gamma$-spacetime.

In other words, the $\gamma$-metric is an appealing candidate to describe space--times close to Schwarzschild, by means of exact analytical solutions to Einstein vacuum equations. This of course is of utmost relevance  and explains why it has been so extensively studied in the past (see \cite{g00, g0, g2, g9, g10, g11, g1, g14, g3, g4, g12, g4b, g4bb, g5, g6, g7, g8, g13, g15, g16} and references therein).

This line of research is further motivated by a  promising  new trend of investigations  aimed to develop tests of gravity theories and corresponding black hole (or naked singularities) solutions  for strong gravitational fields, which is based on  the recent observations of shadow images of the gravitationally collapsed objects at the center of the elliptical galaxy $M87$  and at the center of the Milky Way galaxy by the Event Horizon Telescope (EHT) Collaboration \cite{et1, et2}. The important point is that GR has not been tested yet for such strong fields \cite{et3, et4, et5}. The data from EHT observations can be used to get constraints on the parameters of the mathematical solutions that could describe the geometry surrounding  those  objects. These solutions include, among others,   black hole space--times in modified and alternative theories of gravity \cite{ex1, ex2, ex3, ex4, ex5}, naked singularities as well as classical GR black hole with hair or immersed in matter fields \cite{ex7, ex8, ex9, ex10, ex11, ex12}.

Our purpose in this paper is to provide another yet static non--spherical exact solution  to vacuum Einstein equations, which  could be tested against the results of the  Event Horizon Telescope (EHT) Collaboration. For doing that we shall analyze in detail the geodesics  of test particles in the field of the quasi--hyperbolically $\gamma$ metric.

\section{The $\gamma$-metric and its hyperbolic version}
In Erez--Rosen coordinates the line element  for the $\gamma$-metric is
\begin{equation} \label{gamma}
ds^2=fdt^2-f^{-1}[gdr^2+hd\theta^2+(r^2-2mr)\sin^2\theta d\phi^2],
\end{equation}
where
\begin{eqnarray}
f&=&\left(1-\frac{2m}{r}\right)^\gamma,\\
g&=&\left(\frac{1-\frac{2m}{r}}{1-\frac{2m}{r}+\frac{m^2}{r^2}\sin^2\theta}\right)^{\gamma^2-1},\\
h&=&\frac{r^2\left(1-\frac{2m}{r}\right)^{\gamma^2}}{\left(1-\frac{2m}{r}+\frac{m^2}{r^2}\sin^2{\theta}\right)^{\gamma^2 -1}},
\end{eqnarray}
and $\gamma$ is a constant parameter.

The mass (monopole) $M$  and the quadrupole moment  $Q$ of the solution are given by

\begin{equation}
M=\gamma m, \qquad Q=\gamma (1-\gamma^2)\frac{m^3}{3},
\label{mq}
\end{equation}
implying that the source will be oblate (prolate) for $\gamma>1$ ($\gamma<1$). Obviously for $\gamma=1$ we recover the Schwarzschild solution.

The hyperbolic version of (\ref{gamma}) reads
\begin{equation}\label{gammah}
ds^2=Fdt^2-F^{-1}[Gdr^2+Hd\theta^2+(2mr-r^2)\sinh^2\theta d\phi^2],
\end{equation}
where
\begin{eqnarray}
F&=&\left(\frac{2m}{r}-1\right)^\gamma,\\
G&=&\left(\frac{\frac{2m}{r}-1}{\frac{2m}{r}-1+\frac{m^2}{r^2}\sinh^2\theta}\right)^{\gamma^2-1},\\
H&=&\frac{r^2\left(\frac{2m}{r}-1\right)^{\gamma^2}} {\left(\frac{2m}{r}-1+\frac{m^2}{r^2}\sinh^2\theta\right)^{\gamma^2-1}},
\end{eqnarray}
which can be very easily obtained by following the procedure used in \cite{1} to obtain (\ref{w3}) from (\ref{w2}). It is easy to check that (\ref{gammah}) is a solution to vacuum Einstein equations, and that  $\gamma=1$ corresponds to the line element (\ref{w3}).

Thus, as in \cite{1}, we shall assume that the line element defined by (\ref{gammah}) describes the region $r<2m$, whereas the space--time outside $r=2m$ is described by the ``usual'' $\gamma$-metric (\ref{gamma}). However in this case if $\gamma\neq 1$ the surface $r=2m$ represents a naked singularity since the curvature invariants are singular on that surface (as expected from the Israel theorem \cite{Israel}).

Indeed, the calculation of the Kretschmann scalar $\mathcal{K}$ 
\begin{equation}
\mathcal{K}=R_{\alpha\beta\mu\nu}R^{\alpha\beta\mu\nu},
\end{equation}
for (\ref{gammah}) produces
\begin{eqnarray}
\mathcal{K}&=&\frac{64m^2\gamma^2 (\frac{2m}{r}-1)^{2\gamma}(1+\frac{m^2 \sinh^2\theta}{2mr-r^2})^{2\gamma^2}}{r^2(-2m+r)^2(-m^2+4mr-2r^2+m^2\cosh 2\theta)^3}
\nonumber \\&&\left \{-6r^4+12mr^3(2+\gamma)+3m^3 r(1+\gamma)^2(4+\gamma) \right.\nonumber\\
&-&\left. m^4(1+\gamma)^2 (1+\gamma+\gamma^2)(1-\cosh2\theta)\right. \nonumber \\ &&\left. -3m^2r^2[10+3\gamma (4+\gamma)]+m^2[3r^2\gamma^2\right. \\ &&\left.-3mr\gamma(1+\gamma)^2]\cosh2\theta\right \},
\end{eqnarray}

which is singular at $r=2m$, except for $\gamma=1$, in which case we get
\begin{equation}
\mathcal{K}=\frac{48m^2}{r^6}.
\end{equation}

As it is evident the metric (\ref{gammah}) does not admit the three Killing vectors (\ref{2cmhy}), as well as the $\gamma$-metric (\ref{gamma}) does not admit the Killing vectors (\ref{2cmh}) describing the spherical symmetry.

Indeed,  from
\begin{equation}
\mathcal{L}_\xi g_{\alpha \beta} =\xi^\rho \partial_\rho g_{\alpha \beta} +g_{\alpha \rho}\partial_\beta \xi^\rho+g_{\beta \rho}\partial_\alpha \xi^\rho,
\label{mod1}
\end{equation}
where $\mathcal{L}_\xi $ denotes the Lie derivative with respect to the vectors (\ref{2cmh}),  we obtain  for (\ref{gamma}) two non--vanishing independent components of (\ref{mod1})
\begin{equation}
  \mathcal{L}_\xi g_{\alpha \beta} K^\alpha K^\beta= \mathcal{L}_\xi g_{\alpha \beta} L^\alpha L^\beta=\frac{m^2(1-\gamma^2)\sin 2\theta \cos\phi}{r^2\left(1-\frac{2m}{r}+\frac{m^2}{r^2} \sin^2 \theta\right)},
\end{equation}

\begin{eqnarray}
  \mathcal{L}_\xi {g}_{\alpha \beta} {L}^\alpha {S}^\beta=-\frac{\sin \phi}{\sin \theta} \left [  \left (\frac{1-\frac{2m}{r}}{1-\frac{2m}{r}+\frac{m^2 }{r^2}\sin^2\theta}\right )^{\frac{\gamma^2-1}{2}}  \right.    \\ \left. - \left (\frac{1-\frac{2m}{r}+\frac{m^2 }{r^2}\sin^2\theta}{1-\frac{2m}{r}}\right )^{\frac{\gamma^2-1}{2}}  \right ]\nonumber,
\end{eqnarray}

where the orthogonal tetrad associated to (\ref{gamma}) is
\begin{eqnarray}
V^\alpha&=&\left(\frac{1}{\sqrt{f}}, 0, 0, 0\right), \quad K^\alpha=\left(0, \sqrt{f/g},  0, 0\right), \nonumber \\ L^\alpha&= &\left(0, 0, \sqrt{f/h}, 0\right),\quad S^\alpha=\left(0, 0, 0, \frac{\sqrt{f}}{r \sin\theta\sqrt{1-\frac{2m}{r}}}\right).\nonumber
\end{eqnarray}

On the other hand calculating $\mathcal{L}_\chi g_{\alpha \beta}$ for (\ref{gammah})  and (\ref{2cmhy}) we obtain two non--vanishing components
\begin{equation}
  \mathcal{L}_\chi  {g}_{\alpha \beta} \tilde{K}^\alpha \tilde{K}^\beta= \mathcal{L}_\chi \tilde{g}_{\alpha \beta} \tilde{L}^\alpha \tilde{L}^\beta=\frac{m^2(\gamma^2-1)\sinh 2\theta \cos \phi}{r^2\left(\frac{2m}{r}-1+\frac{m^2}{r^2} \sinh^2 \theta\right)},
\end{equation}

\begin{eqnarray}
  \mathcal{L}_\chi {g}_{\alpha \beta} \tilde{L}^\alpha \tilde{S}^\beta=\frac{\sin \phi}{\sinh \theta} \left [ \left (\frac{\frac{2m}{r}-1}{\frac{2m}{r}-1+\frac{m^2 }{r^2}\sinh^2\theta}\right )^{\frac{\gamma^2-1}{2}}\right.  \\ \left.- \left (\frac{\frac{2m}{r}-1+\frac{m^2 }{r^2}\sinh^2\theta}{\frac{2m}{r}-1}\right )^{\frac{\gamma^2-1}{2}}  \right ]\nonumber,
\end{eqnarray}
where the orthogonal tetrad associated to (\ref{gammah}) is

\begin{eqnarray}
\tilde V^\alpha&=&\left(\frac{1}{\sqrt{F}}, 0, 0, 0\right), \quad \tilde K^\alpha=\left(0, \sqrt{F/G},  0, 0\right), \nonumber \\ \tilde L^\alpha&= &\left(0, 0, \sqrt{F/H}, 0\right),\quad \tilde S^\alpha=\left(0, 0, 0, \frac{\sqrt{F}}{r \sinh\theta\sqrt{\frac{2m}{r}-1}}\right).\nonumber
\end{eqnarray}

In other words the $\gamma$-metric deviates from spherical symmetry in a similar way as the hyperbolic version of the $\gamma$-metric deviates from hyperbolical symmetry. This is  the origin of the term ``quasi--hyperbolically symmetric'' applied to (\ref{gammah}).

 \section{Geodesics}
We shall now find the geodesic equations for test particles in the
metric (\ref{gammah}).  The qualitative differences in the trajectories of the test
particles
as compared with the $\gamma$- metric and  the metric (\ref{w3}) will be brought out and discussed.

The equations governing the geodesics can be derived from the Lagrangian
\begin{equation}\label{lag1}
2{\cal L}=g_{\alpha\beta}\dot{x}^\alpha\dot{x}^\beta,
\end{equation}
where the dot denotes differentiation with respect to an affine parameter
$s$ , which for timelike geodesics coincides with the proper time.

Then,
from the Euler-Lagrange equations,
\begin{equation}\label{lag3}
\frac{d}{ds}\left(\frac{\partial{\cal
L}}{\partial\dot{x}^\alpha}\right)-\frac{\partial{\cal L}}
{\partial x^\alpha}=0,
\end{equation}

\noindent we obtain for  (\ref{gammah})

\begin{equation}\label{geo111}
  \ddot{t}-\frac{2\gamma m }{r^2(\frac{2m}{r}-1)}\dot{r}\dot{t} =0,
\end{equation}

\begin{eqnarray}\label{ddotr}
  \ddot{r}-\frac{m\gamma (\frac{2m}{r}-1)^{2\gamma-\gamma^2}}{r^2(\frac{2m}{r}-1+\frac{m^2 }{r^2}\sinh^2\theta)^{1-\gamma^2}}\dot{t}^2\nonumber \\
-\frac{m}{r^2}\left [\frac{(\gamma^2-\gamma-1)}{\frac{2m}{r}-1}-\frac{(\gamma^2-1)\left(1+\frac{m}{r}\sinh^2{\theta}\right)}{\frac{2m}{r}-1+\frac{m^2}{r^2}\sinh^2{\theta}} \right ]\dot{r}^2\nonumber
  \\
 - \frac{m^2 (\gamma^2-1)\sinh 2\theta}{r^2 \left(\frac{2m}{r}-1+\frac{m^2}{r^2}\sinh^2{\theta}\right)}\dot{\theta}\dot{r}+\left [r+m(\gamma^2-\gamma-2)\right. \nonumber \\ \left.-\frac{m (\gamma^2-1)(1+\frac{m}{r}\sinh^2{\theta})\left(\frac{2m}{r}-1\right)}{\left(\frac{2m}{r}-1+\frac{m^2}{r^2}\sinh^2{\theta}\right)}\right ]\dot{\theta}^2 \nonumber \\
 - \frac{[m (1+\gamma)-r](\frac{2m}{r}-1+\frac{m^2}{r^2}\sinh^2\theta)^{\gamma^2-1}\sinh^2\theta}{(\frac{2m}{r}-1)^{\gamma^2-1}}\dot{\phi}^2 \nonumber\\=0,\nonumber \\
\end{eqnarray}

\begin{eqnarray}\label{dottheta}
  &&\ddot{\theta} +  \frac{m^2 (\gamma^2-1)\sinh 2\theta}{2r^4 \left(\frac{2m}{r}-1\right)\left(\frac{2m}{r}-1+\frac{m^2}{r^2}\sinh^2{\theta}\right)} \dot{r}^2\nonumber \\&+& 2\left[\frac{1}{r}+\frac{m(\gamma-\gamma^2)}{r^2 \left(\frac{2m}{r}-1\right)}+ \frac{m (\gamma^2-1)\left(1+\frac{m}{r}\sinh^2{\theta}\right)} {r^2 \left(\frac{2m}{r}-1+\frac{m^2}{r^2}\sinh^2{\theta}\right)}\right]\dot{\theta}\dot{r}\nonumber\\
  &-&\frac{m^2(\gamma^2-1)\sinh 2\theta}{2r^2 \left(\frac{2m}{r}-1+\frac{m^2}{r^2}\sinh^2{\theta}\right)}\dot{\theta}^2 \nonumber  \\&-&
  \frac{(\frac{2m}{r}-1)^{1-\gamma^2}\sinh 2\theta
  }{2(\frac{2m}{r}-1+\frac{m^2}{r^2}\sinh^2\theta)^{1-\gamma^2}}\dot{\phi}^2=0,
\end{eqnarray}

\begin{equation}\label{geo44}
  \ddot{\phi}+\frac{2}{r^2 \left(\frac{2m}{r}-1\right)}[m(1+\gamma)-r]\dot{r} \dot{\phi}+(2\coth{\theta} )\dot{\theta}\dot{\phi}=0.
\end{equation}

Let us first analyze some particular cases, from  which  some important general results on the geodesic structure of the system may be deduced.

Thus, let us assume that at some given initial $s=s_0$ we have $\dot \theta=0$, then it follows at once from (\ref{dottheta}) that such a condition will propagate in time only if $\theta=0$. In other words, any $\theta=constant$ trajectory is unstable except $\theta=0$. It is worth stressing the difference between this case and the situation in the purely hyperbolic metric  where  $\dot \phi=0$ also ensures stability.

Next, let us consider the case of circular orbits. These are defined by  $\dot{r}=\dot{\theta}=0$, producing

\begin{eqnarray}
\ddot{t} = \ddot{\phi}&=&0 ,\\
  m\gamma \dot{t}^2+\frac{r^2\left [(\gamma+1)m-r   \right ]}{(\frac{2m}{r}-1)^{2\gamma-1}}\sinh^2\theta\dot{\phi}^2 &=& 0 ,\\
  \sinh\theta \cosh \theta \dot{\phi}^2&=& 0\label{cg}.
\end{eqnarray}

From (\ref{cg}) it is obvious that, as for the hyperbolically symmetric black hole, no circular geodesics exist in this case,  which is  at variance with the $\gamma$-metric space--time.

Let us now consider the motion of a test particle along a meridional
line $\theta$ ($\dot r=\dot \phi=0$). In this case as shown in \cite{2nc} motion is forbidden if $\gamma=1$, however from (\ref{ddotr}) it is a simple matter to see that for $\gamma>1$ there are possible solutions.

More so,  let us assume (always in the purely meridional motion case) that at $s=0$ we have $\theta=constant\neq 0$ and
$\dot{\theta}=0$. Then, if $\gamma=1$, it follows from (\ref{dottheta}) that
$\ddot{\theta}=0$. The particle remains on the same plane, a result already obtained in \cite{2nc}.
However if $\gamma\neq 1$,  $\ddot{\theta}$ does not
need to vanish, and the particle leaves the  plane ($\theta=constant$).

This effect implies the existence of a force parallel to the axis of symmetry, a result similar to the one obtained for the $\gamma$-metric, and which illustrates
further the influence of  the deviation from the hyperbolically symmetric case.

Let us consider the case of purely radial geodesics described by $\dot \theta=\dot \phi=0$, producing
\begin{equation}\label{geo11}
  \ddot{t}-\frac{2\gamma m }{r^2(\frac{2m}{r}-1)}\dot{r}\dot{t} =0,
\end{equation}

\begin{eqnarray}\label{dotr}
  \ddot{r}-\frac{m\gamma (\frac{2m}{r}-1)^{2\gamma-\gamma^2}}{r^2(\frac{2m}{r}-1+\frac{m^2 }{r^2}\sinh^2\theta)^{1-\gamma^2}}\dot{t}^2\nonumber \\
-\frac{m}{r^2}\left [\frac{(\gamma^2-\gamma-1)}{\frac{2m}{r}-1}-\frac{(\gamma^2-1)\left(1+\frac{m}{r}\sinh^2{\theta}\right)}{\frac{2m}{r}-1+\frac{m^2}{r^2}\sinh^2{\theta}} \right ]\dot{r}^2=0,\nonumber\\
\end{eqnarray}

\begin{equation}
 \frac{m^2 (\gamma^2-1)\sinh 2\theta}{2r^4 \left(\frac{2m}{r}-1\right)\left(\frac{2m}{r}-1+\frac{m^2}{r^2}\sinh^2{\theta}\right)} \dot{r}^2=0.
\end{equation}

The last of the equations  above indicates that, if $\gamma\neq 1$, purely radial geodesics only exists along the axis $\theta=0$.

In this case  it follows from (\ref{lag3}), due to the symmetry imposed
\begin{equation}\label{lag4}
\frac{\partial{\cal L}}{\partial\dot{t}}=\mbox{constant}=E =
\dot{t}\left(\frac{2m}{r}-1\right)^\gamma,
\end{equation}
\begin{equation}
\frac{\partial{\cal L}}{\partial\dot{\phi}}=\mbox{constant}=L =
-\dot{\phi}\left(\frac{2m}{r}-1\right)^{1-\gamma}r^2 \sinh^2\theta,
\end{equation}
where $E$ and $L$ represent, respectively, the total energy and the angular
momentum of the test particle.  Since we have already seen that the only stable radial trajectory  is $\theta=0$ the angular momentum vanishes for those trajectories.

Then using (\ref{lag4}) we obtain  for the first integral of (\ref{dotr})
\begin{equation}
\dot{r}^2=E^2-V^2,
\end{equation}
where $V$, which can be associated to the potential energy of the test
particle, is given by
\begin{equation}\label{vtetaceo}
V^2\equiv\left(\frac{2m}{r}-1\right)^\gamma,
\end{equation}

or, introducing the
dimensionless variable $x\equiv r/m$,  (\ref{vtetaceo}) ) becomes
\begin{equation}
V^2=\left(\frac{2}{x}-1\right)^\gamma.
\end{equation}

As we see from Fig. 1, for any given  value of $E$ (however large, but finite), the test particle inside the naked singularity  never reaches the center, moving between the closest point to the center where $E=V$, and $x=\infty$  since  nothing prevents the particle to  cross the naked singularity outwardly. It is possible however, since  for $x>2$ the space--time is no longer  described  by (\ref{gammah}) but by the usual $\gamma$-metric (\ref{gamma}), that for some value of $E$ the particle bounces back  at a  point ( $x>2$) where $E=V$.

Thus for this particular value of energy we have a bounded trajectory with extreme points at both sides of the naked singularity. For sufficiently large (but finite) values of energy, the trajectory is unbounded and  the particle moves between a point close to, but at finite distance from, the center and $r\rightarrow \infty$.

The above  picture  is quite different from the behavior of the test particle in the $\gamma$-metric as described  in \cite{gammas}, and 
similar to the one observed for a radially moving test particle inside the horizon for the metric (\ref{w3}).  However, in our case, the parameter $\gamma$ affects the behavior of the test particle as it is apparent from Figure 1. Specifically, for $\gamma>1$ the test particle is repelled stronger from the center, bouncing back  at values of $r$ larger than in the case $\gamma \leq 1$.

\begin{figure}
  \centering
  \includegraphics[width=5.5cm]{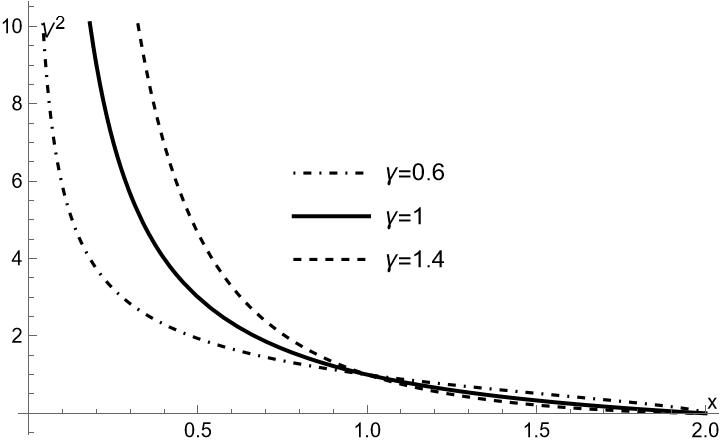}
  \caption{$V^2$ as function of  $x$, for the three values of $\gamma$ indicated in the figure}
\end{figure}

In order to understand the results above, it is convenient to calculate  the four--acceleration of  a static observer in the frame of (\ref{gammah}). We recall that a static observer is one whose four velocity $U^\mu$ is proportional to the Killing time--like vector \cite{Caroll}, i.e.
\begin{equation}
U^\mu=\left[\frac{1}{(\frac{2m}{r}-1)^{\gamma/2}}, 0, 0, 0 \right].
\label{1a}
\end{equation}
Then for the four--acceleration $a^\mu\equiv U^\beta U^\mu_{;\beta}$ we obtain for the region inner to the naked singularity

\begin{equation}\label{3a}
a^\mu=\left[0,-\frac{m \gamma\left(\frac{2m}{r}-1+\frac{m^2}{r^2}\sinh^2{\theta}\right)^{\gamma^2 -1}}{r^2\left(\frac{2m}{r}-1\right)^{\gamma^2-\gamma}}, 0, 0\right],
\end{equation}
whereas for the region outside the naked singularity, described by  (\ref{gamma})  we obtain, with
\begin{equation}
U^\mu=\left[\frac{1}{(1-\frac{2m}{r})^{\gamma/2}}, 0, 0, 0 \right],
\label{1a}
\end{equation}

\begin{equation}\label{3ab}
a^\mu=\left[0,\frac{m \gamma\left(1-\frac{2m}{r}+\frac{m^2}{r^2}\sin^2{\theta}\right)^{\gamma^2 -1}}{r^2\left(1-\frac{2m}{r}\right)^{\gamma^2-\gamma}}, 0, 0\right].
\end{equation}

The physical meaning of (\ref{3a}) and (\ref{3ab}) is clear, it  represents  the inertial radial acceleration, which is necessary in order to maintain static the frame, by canceling the gravitational acceleration exerted on the frame, for the space--times (\ref{gammah}) and (\ref{gamma}) respectively. Since this acceleration is directed radially inwardly (outwardly), in the region inner (outer) to the naked singularity,  it means that the gravitational force is repulsive (attractive). The attractive nature of gravitation in (\ref{gamma}) is expected, whereas its repulsive nature in (\ref{gammah}) is characteristic of hyperbolical space--times, and  explains the peculiarities of the orbits inside the horizon. In particular, we see from (\ref{3a}) that the absolute value of the radial acceleration grows with $\gamma$, implying that the repulsion is stronger for larger $\gamma$, as it follows from the Figure 1.

We shall next consider the geodesics in the $\theta-r$ plane ($\phi=constant$). The interest of this case becomes intelligible if we recall that our space--time (\ref{gammah}) is axially symmetric, implying that the general properties of motion on any slice $\phi=constant$ would be invariant with respect to rotation around the symmetry axis.

In this case, geodesic equations read

\begin{equation}\label{geo11}
 \ddot{t}-\frac{2\gamma m }{r^2(\frac{2m}{r}-1)}\dot{r}\dot{t} =0,
  \end{equation}
\begin{eqnarray}
  \ddot{r}-\frac{m\gamma (\frac{2m}{r}-1)^{2\gamma-\gamma^2}}{r^2(\frac{2m}{r}-1+\frac{m^2 }{r^2}\sinh^2\theta)^{1-\gamma^2}}\dot{t}^2\nonumber \\
-\frac{m}{r^2}\left [\frac{(\gamma^2-\gamma-1)}{\frac{2m}{r}-1}-\frac{(\gamma^2-1)\left(1+\frac{m}{r}\sinh^2{\theta}\right)}{\frac{2m}{r}-1+\frac{m^2}{r^2}\sinh^2{\theta}} \right ]\dot{r}^2\nonumber
  \\
 - \frac{m^2 (\gamma^2-1)\sinh 2\theta}{r^2 \left(\frac{2m}{r}-1+\frac{m^2}{r^2}\sinh^2{\theta}\right)}\dot{\theta}\dot{r}+\left [r+m(\gamma^2-\gamma-2)\right. \nonumber \\ \left.-\frac{m (\gamma^2-1)(1+\frac{m}{r}\sinh^2{\theta})\left(\frac{2m}{r}-1\right)}{\left(\frac{2m}{r}-1+\frac{m^2}{r^2}\sinh^2{\theta}\right)}\right ]\dot{\theta}^2 =0,\nonumber \\
\end{eqnarray}
\begin{eqnarray}\label{ndt}
 && \ddot{\theta} +  \frac{m^2 (\gamma^2-1)\sinh 2\theta}{2r^4 \left(\frac{2m}{r}-1\right)\left(\frac{2m}{r}-1+\frac{m^2}{r^2}\sinh^2{\theta}\right)} \dot{r}^2\nonumber \\&+& 2\left[\frac{1}{r}+\frac{m(\gamma-\gamma^2)}{r^2 \left(\frac{2m}{r}-1\right)}+ \frac{m (\gamma^2-1)\left(1+\frac{m}{r}\sinh^2{\theta}\right)} {r^2 \left(\frac{2m}{r}-1+\frac{m^2}{r^2}\sinh^2{\theta}\right)}\right]\dot{\theta}\dot{r}\nonumber\\
  &-&\frac{m^2(\gamma^2-1)\sinh 2\theta}{2r^2 \left(\frac{2m}{r}-1+\frac{m^2}{r^2}\sinh^2{\theta}\right)}\dot{\theta}^2 =0.
\end{eqnarray}

\noindent To simplify the calculations we shall adopt a perturbative approach assuming  $\gamma=1+\epsilon$, for $\epsilon<<1$, and neglecting terms of order   $\epsilon ^2$  and higher. Doing so we obtain from  (\ref{ndt}) at order $O(0)$ and $O(\epsilon)$ respectively
\begin{equation}\label{dottheta0}
  (\dot{\theta}r^2\dot{)}=0\quad \Rightarrow \quad \dot{\theta}=\frac{c_1}{r^2},
\end{equation}
and 
\begin{eqnarray}
 && \frac{m^2\sinh 2\theta \, \dot{r}^2}{r^4(\frac{2m}{r}-1)(\frac{2m}{r}-1+\frac{m^2}{r^2}\sinh ^2\theta)}\nonumber \\&-&\frac{m^2 \sinh 2\theta \, \dot{\theta}^2}{r^2(\frac{2m}{r}-1+\frac{m^2}{r^2}\sinh^2\theta)}\nonumber\\
  &-&\frac{2m}{r^2}\left [ \frac{1-\frac{2m}{r}+(\frac{2m}{r}-\frac{3m^2}{r^2})\sinh^2\theta}{(\frac{2m}{r}-1)(\frac{2m}{r}-1+\frac{m^2}{r^2}\sinh ^2\theta)} \right ]\dot{\theta}\,\dot{r}=0.\nonumber \\ \label{dottheta1}
\end{eqnarray}

\noindent   Introducing
\begin{equation}
\dot{r}=r_\theta \dot {\theta},\qquad y=\frac{m}{r},
\end{equation}
equation  (\ref{dottheta1}) becomes

\begin{equation}
y^2_\theta+\frac{2y_\theta}{\sinh 2\theta}\left[1-2y+(2y-3y^2)\sinh ^2\theta \right]-y^2(2y-1)=0,  \label{dottheta2}
\end{equation}
whose integration produces
\begin{equation}
y=constant=1/2.
\label{per}
\end{equation}

\noindent The order $O(0)$  can be easily calculated from (\ref{ddotr}) and (\ref{dottheta0}), producing
\begin{equation}\label{ddotr01}
  \ddot{r}-\frac{m E^2}{r^2(\frac{2m}{r}-1)}+\frac{m \, \dot{r}^2}{r^2(\frac{2m}{r}-1)}+\frac{(r-2m)c_1^2}{r^4}=0,
\end{equation}

\noindent whose first integral reads

\begin{equation}
\dot{r}=\sqrt{E^2-(\frac{2m}{r}-1)(\frac{c_1^2}{r^2}+1)},\label{pint}
\end{equation}

\noindent   or, introducing the variable $z$
\begin{equation}
\dot{r}=r_\theta \dot {\theta},\qquad z\equiv 2y=\frac{2m}{r},
\end{equation}

\begin{equation}
z_\theta=\frac{1}{k}\sqrt{E^2-(z-1)(k^2z^2+1)},
\end{equation}

\noindent with  $c_1=-2mk$.

\noindent This equation was already obtained and solved for the case $\gamma=1$ (eq.  (38) in \cite{2nc}), with the boundary condition that all trajectories coincide at $\theta =0, z=1$. Here we present the integration of such an equation for the values indicated in the Figure 2 (please notice that we have used for this figure the variable $z=2y$ in order to keep the same notation for the order $O(0)$ as in \cite{2nc}). 

Let us now analyze in some detail the physical implications of Figure 2 and equation (\ref{per}).  As we see the solution of the order $0(\epsilon)$  maintains  a constant value of $y$ which is the same value assumed in the boundary condition. At order $O(0)$ we see from Figure 2, that the particle never reaches the center, which may only happen as $k$ and $E$ tend to infinity. In either case the particle never crosses outwardly the surface  $y=1/2$ ($z=1$), this happening only along the radial geodesic $\theta=0$. The influence of $\gamma$ in the final picture can be deduced by combining Figure  2 and equation (\ref{per}).

\begin{figure}
  \centering
  \includegraphics[width=5.5cm]{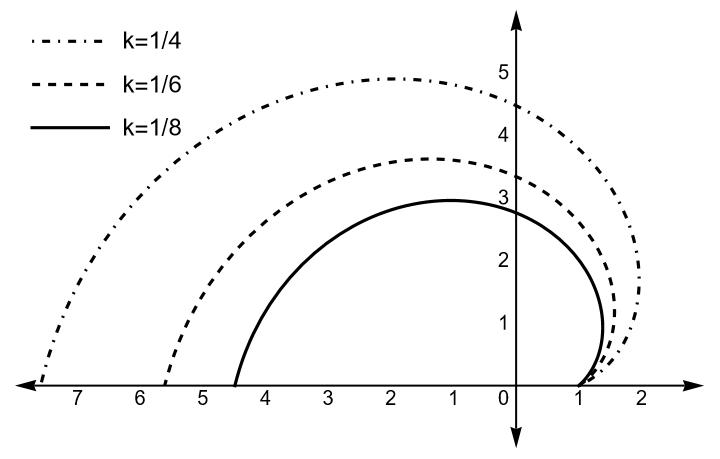}
  \caption{ $z\equiv\frac{2m}{r} $ as function of  $\theta$, for the values of $k$ indicated on the figure and $E=3$.}
\end{figure}
\section{Conclusions}

Motivated by the relevance of the $\gamma$-metric (\ref{gamma}) and the hyperbolically symmetric metric (\ref{w3}), we have proposed in this work to analyze the physical properties of the hyperbolical version of the $\gamma$-metric. Such space--time described by the line element (\ref{gammah}) shares with the hyperbolically symmetric space--time described by (\ref{w3}) some important features, the most relevant of which is the repulsive character of gravity inside the surface $r=2m$. On the other hand, as for the $\gamma$-metric (\ref{gamma}), the surface $r=2m$ is not regular, thereby describing a naked singularity. The  space--time (\ref{gammah}) is not hyperbolically symmetric in the sense that it does not admit the Killing vectors (\ref{2cmhy}), a fact suggesting the name ``quasi--hyperbolically symmetric''  for such space--time. 

We have focused our study on  the characteristics  of the motion of test particles in the space--time described by (\ref{gammah}), with special attention payed on the role of the parameter $\gamma$. Thus our main conclusions are:
\begin{enumerate}
\item Test particles may cross the surface $r=2m$ outwardly, but only along the axe $\theta=0$. This situation appears in the study  of the geodesics in (\ref{w3}) presented in \cite{2nc}, however in our case the  distinctive repulsive force  of this space--time is increased by the parameter $\gamma$.
\item Like in the hyperbolically symmetric case, the test particles never reach the center, however in our case the test particles radially directed to the center  bounce back farther from the center as $\gamma$ increases. This result becomes intelligible from a simple inspection of  (\ref{3a}).
\item The motion of test particles on any slice $\phi=constant$ though  qualitatively similar to the case $\gamma=1$, is affected by the value of $\gamma$ as follows from the analysis of Figure 2 and (\ref{per}).
\end{enumerate}

As we mentioned before, a new line of investigations based on observations of shadow images of the gravitationally collapsed aiming to 
the tests of gravity theories and corresponding black hole (or naked singularities) solutions  for strong gravitational fields, is right now attracting the interest of many researchers. Such studies are particularly  suitable  for contrasting the physical relevance of different exact solutions to the field equations. We believe that the metric here exhibited deserves to be considered as a suitable  candidate for such comparative studies.  However it is worth mentioning that we have restricted our study to time--like geodesics, whereas any contrast with ETH observational data, would require  results obtained from the study of null geodesics. Notwithstanding, the results obtained for time--like geodesics here presented, point to the potential of the metric under consideration.

We would like to conclude with a mention to what we believe is one of the most promising application of hyperbolical metrics. We have in mind the modeling of 
 extragalactic relativistic jets.  It should be clear that at  present, such an application remains within the realm of speculation, however the comments below justify our (moderate) optimism.
 
Relativistic jets are highly energetic phenomena which have been observed in many systems (see \cite{Bladford, Margon, Sams,blan} and references therein), usually associated with the presence of a compact object, and exhibiting a high degree of collimation.  Since no consensus has been reached until now, concerning the basic mechanism explaining these two features of jets (collimation and high energies), we feel motivated to  speculate that the metric here considered   could be considered as a possible engine behind the jets.

 Indeed, on the one hand the collimation is ensured by the fact that test particles may cross the naked singularity outward, but only along the $\theta=0$ axis. On the other hand, as implied by (\ref{3a}), the strength of the repulsive gravitational force acting on the particle as $r\rightarrow 0$ increases as $\frac{1}{r^{(\gamma+2})}$. Explaining the high energies of particles bouncing back from regions close to $r=0$. More so,  the fact  that the repulsive force  would be larger for larger values of $\gamma$,  enhances further  the efficiency  of our model as the engine of such jets, as compared with the $\gamma=1$ case.
 
  It goes without saying that the confirmation of this mechanism requires a much more detailed setup based on astronomical observations of jets, which  is clearly out of the scope of this work.

\begin{acknowledgments}
This  work  was  partially supported by the Spanish  Ministerio de Ciencia, Innovaci\'on, under Research Project No. PID2021-122938NB-I00. L. H. also wishes to thank Universitat de les
 Illes Balears for financial support and hospitality. A.D.P. wishes to thank Universitat de les
 Illes Balears for its hospitality.
\end{acknowledgments}

\end{document}